\documentclass[iop,apj]{emulateapj}
\usepackage{amsmath,amssymb,amstext}

\usepackage[breaklinks,colorlinks,citecolor=blue,linkcolor=magenta]{hyperref} 

\usepackage[all]{hypcap} 

\usepackage{natbib}
\usepackage{graphicx}
\usepackage[caption=false]{subfig}
\usepackage{booktabs}

\begin{document}

\title{The effect of ionic correlations on radiative properties in the solar interior and terrestrial experiments}

\author{Menahem Krief}
\email[e-mail address: ]{menahem.krief@mail.huji.ac.il}
\affiliation{Racah Institute of Physics, 
	The Hebrew University, 
	9190401 Jerusalem, Israel}

\author{Yair Kurzweil}
\affiliation{Physics Department, Nuclear Research Center-Negev, P.O. Box 9001, Beer-Sheva, Israel}

\author{Alexander Feigel}
\affiliation{Racah Institute of Physics, 
	The Hebrew University, 
	9190401 Jerusalem, Israel}
\author{Doron Gazit}
\affiliation{Racah Institute of Physics, 
	The Hebrew University, 
	9190401 Jerusalem, Israel}

\begin{abstract}
Intending to solve the decade old problem of solar opacity, we report substantial photoabsorption uncertainty due to the effect of ion-ion correlations.
By performing detailed opacity calculations of the solar mixture, we find that taking into account the ionic structure changes  the Rosseland opacity near the convection zone by $ \sim 10\% $.
  We also report a $ \sim15\% $ difference in the Rosseland opacity for iron, which was recently measured at the Sandia Z facility, 
where the temperature reached that prevailing in the convection zone boundary while the density is $ 2.5 $ times lower. Finally, we propose a method to measure opacities at solar temperatures and densities that were never reached in the past via laboratory radiation flow experiments, by using plastic foams doped with permilles of dominant photon absorbers in the Sun. The method is advantageous for an experimental study of solar opacities that may lead to a resolution of the solar problem.
\end{abstract}

\maketitle


\section{Introduction}
Radiation transport is of decisive importance in the modeling of a wide variety of high energy density plasmas as those encountered in stellar interior as well as in terrestrial laboratories, such as high power laser and Z-pinch facilities. The radiative opacity is a key quantity describing the coupling between radiation and matter which entails a sophisticated interplay of plasma and atomic physics.

Over the past decade, an outstanding open problem in solar physics has emerged, as solar photospheric abundances of metallic elements have been significantly revised downward (\cite{asplund2009chemical}). Standard solar models  cannot reproduce helioseismic measurements, such as the convection zone radius, the surface helium abundance and the sound speed profile, when using these revised abundances. This gave rise to the \textit{solar composition problem}, motivating a rapid growth of research efforts in the field (\cite{serenelli2009new,Bergemann2014,serenelli2016alive,vinyoles2016new}). The implications of this problem on astrophysical science are very large since stellar evolution theory, which is routinely used to interpret any star or stellar population,  is calibrated on the Sun. Therefore, the knowledge about stars and galaxies is nowhere better than the current understanding of the Sun. In addition, the Sun has been extensively  used for setting constraints on the properties of dark matter candidates (\cite{schlattl1999helioseismological}). 

\begin{figure}[]
	\resizebox{0.5\textwidth}{!}{\includegraphics{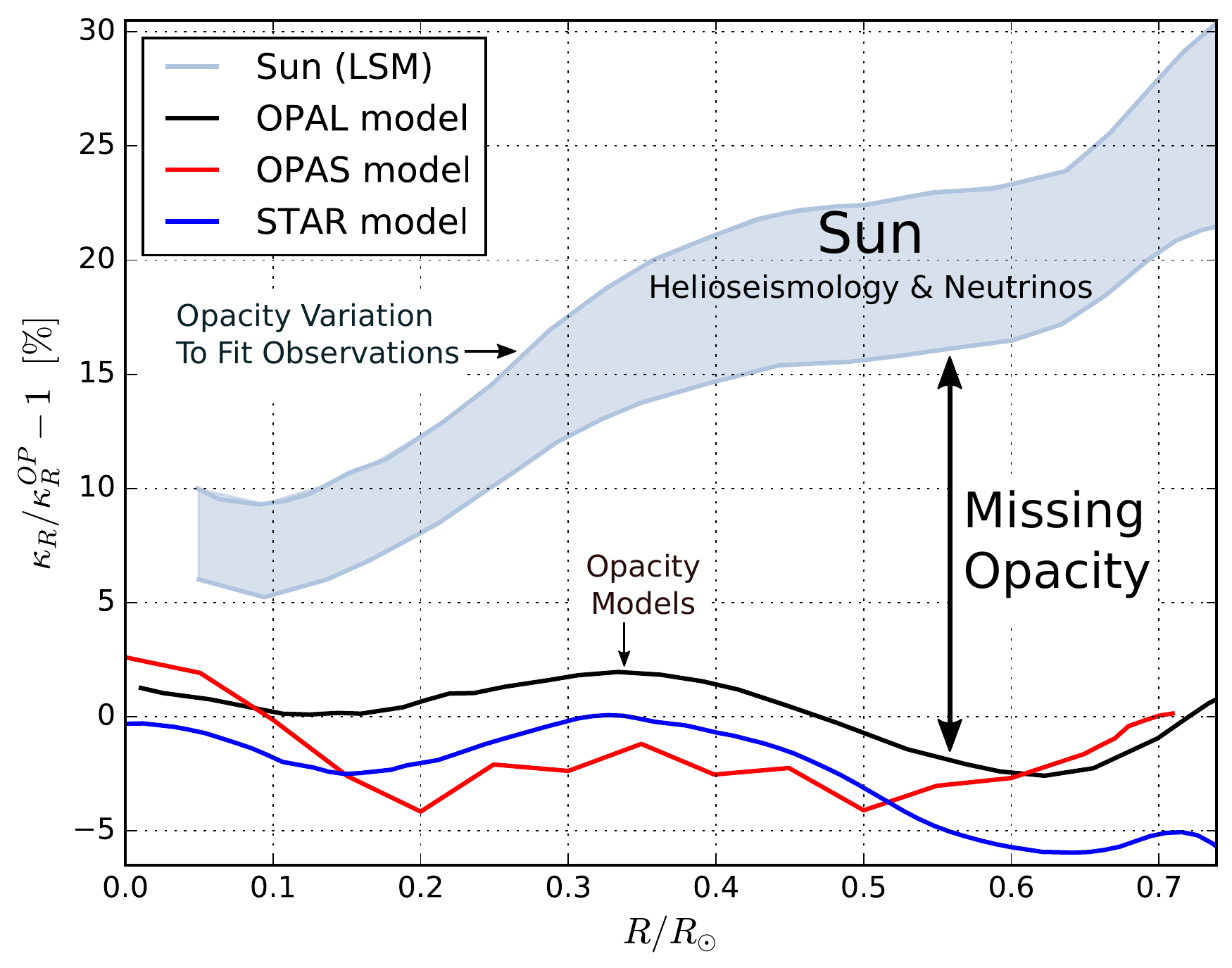}}
	\caption{\textbf{An illustration of the solar opacity problem}.
		The shaded area represents the range of required opacity variation across the solar interior, in order for helioseismic and neutrino observations to agree with solar model predictions. 
		The solid curves represent
		relative differences in the Rosseland opacity $ \kappa_{R}/\kappa^{OP}_{R}-1 $, between the OP opacity model (\cite{seaton1994opacities})
		and the OPAL model (\cite{iglesias1996updated}, in black), the OPAS model (\cite{blancard2012solar,mondet2015opacity}, in red) and the STAR model (\cite{krief2016solar,krief2016line}, in blue). The small discrepancies between opacity models cannot explain the missing opacity that is constrained by observations.
	}
	\label{fig:dk_intro}
\end{figure}

Photon absorption by metallic elements is a major source of opacity in the solar interior. It is widely believed that in order for solar models to  reproduce  helioseismic observations, opacities of mid and high-Z elements, which are the most complicated to model theoretically and for which there is no experimental validation, should be revised upward to compensate for the decreased  abundances.  It has been shown (\cite{christensen2009opacity,villante2014chemical}) that a smooth increase of the solar opacity from $ \sim 5\% $ near the core to $ \sim 20\% $ near the convection zone boundary (CZB), reproduces helioseismic and neutrino observations, as demonstrated in \autoref{fig:dk_intro}. Differences between first principles state of the art opacity models 
(\cite{seaton1994opacities,iglesias1996updated,iglesias2015iron,blancard2012solar,krief2016solar,colgan2016new})
are too small to account for the \textit{missing opacity}, giving rise to a \textit{solar opacity problem} (\cite{serenelli2016implications,vinyoles2017new}). Otherwise, assuming that opacities used currently in solar models are accurate, other parts of solar models must be altered, such as stellar atmosphere and spectroscopic models, mixing mechanisms (\cite{haxton2008cn}) and enhanced gravitational settling, or even non standard solar models (\cite{serenelli2011solar}) incorporating rotation and magnetic fields, which are completely prohibitive with present day computational capabilities.

Recent laboratory measurements (\cite{bailey2015higher}) of the monochromatic opacity of iron, a major photon absorber throughout the solar interior, were performed at the Sandia Z machine. The temperatures reached at these experiments match those found near the CZB, while the electron densities are about a factor of $ 2.5 $ lower. The measured spectra is larger than calculations of all available atomic models by $ 30\%-400\% $. No satisfactory explanation for this discrepancy is yet known (\cite{blancard2016comment}). The observed upward revision of iron opacity supports the revised solar abundances and the missing opacity in solar physics, although at different plasma conditions than those found at the CZB.

Taking into account the complicated many-body interaction  of a photon emitting/absorbing ion with the adjacent hot dense plasma in atomic calculations is a very important yet extremely challenging task in the field.  Opacity models frequently approximate the ionic structure in the framework of the rather crude Ion-Sphere (IS) model (\cite{liberman1979self}), in which the ionic structure is completely smeared out. The sensitivity of solar opacity calculations to the IS approximation was never analyzed in the context of the solar opacity problem.

In this work, plasma correlation effects are pointed out as a significant source of uncertainty in the calculation of solar opacities. The effects of plasma correlations beyond the IS model are studied by using the Ion-Correlation (IC) model by Rozsnyai (\cite{rozsnyai1991photoabsorption,rozsnyai1992solar,rozsnyai2014equation}).
 Detailed opacity calculations of the solar mixture 
  were preformed in order to  compare the IS and IC models. We report a $ \sim 10\% $ increase in the Rosseland opacity near the CZB.
  The effect was also examined at the conditions of the recent Sandia experiment.
Finally, we highlight and analyze the possibility to measure Rosseland opacities via terrestrial radiation flow experiments.
The range of temperature and density that can be achieved in such experiments, if performed at the NIF (\cite{moore2015characterization}), cover the conditions of the solar plasma at solar radii  ranging from the CZB ($ R/R_{\odot}=0.72 $) and deeper, up to $ R/R_{\odot}\approx 0.55 $, which were never achieved in laboratory experiments. 
These experiments are  promising candidates for an experimental study of Rosseland opacities, and can be used to infer directly the uncertainty in calculations of Rosseland opacities at stellar interior conditions, and most importantly, may finally point out weather the use of inaccurate opacities in solar models are the source of the solar abundance problem.

\section{The effect of ion correlation in the solar radiative zone}

The complicated interactions between an ion and its environment give rise to important phenomena such as collisional line broadening (\cite{dimitrijevic1987simple,krief2016line}) and ionization potential depression (\cite{ciricosta2016measurements}). In addition, the plasma environment affects the local atomic potential and the boundary conditions of atomic wavefunctions. In order to calculate the latter effect properly, the ionic structure of the plasma, which is given by the ion-ion pair correlation function must be known. However, it is very challenging to calculate ionic correlations from first principles (\cite{daligault2016ionic}) and approximate models, which were never tested experimentally at stellar interior conditions, must be used.

\begin{figure}[]
	\resizebox{0.5\textwidth}{!}{\includegraphics{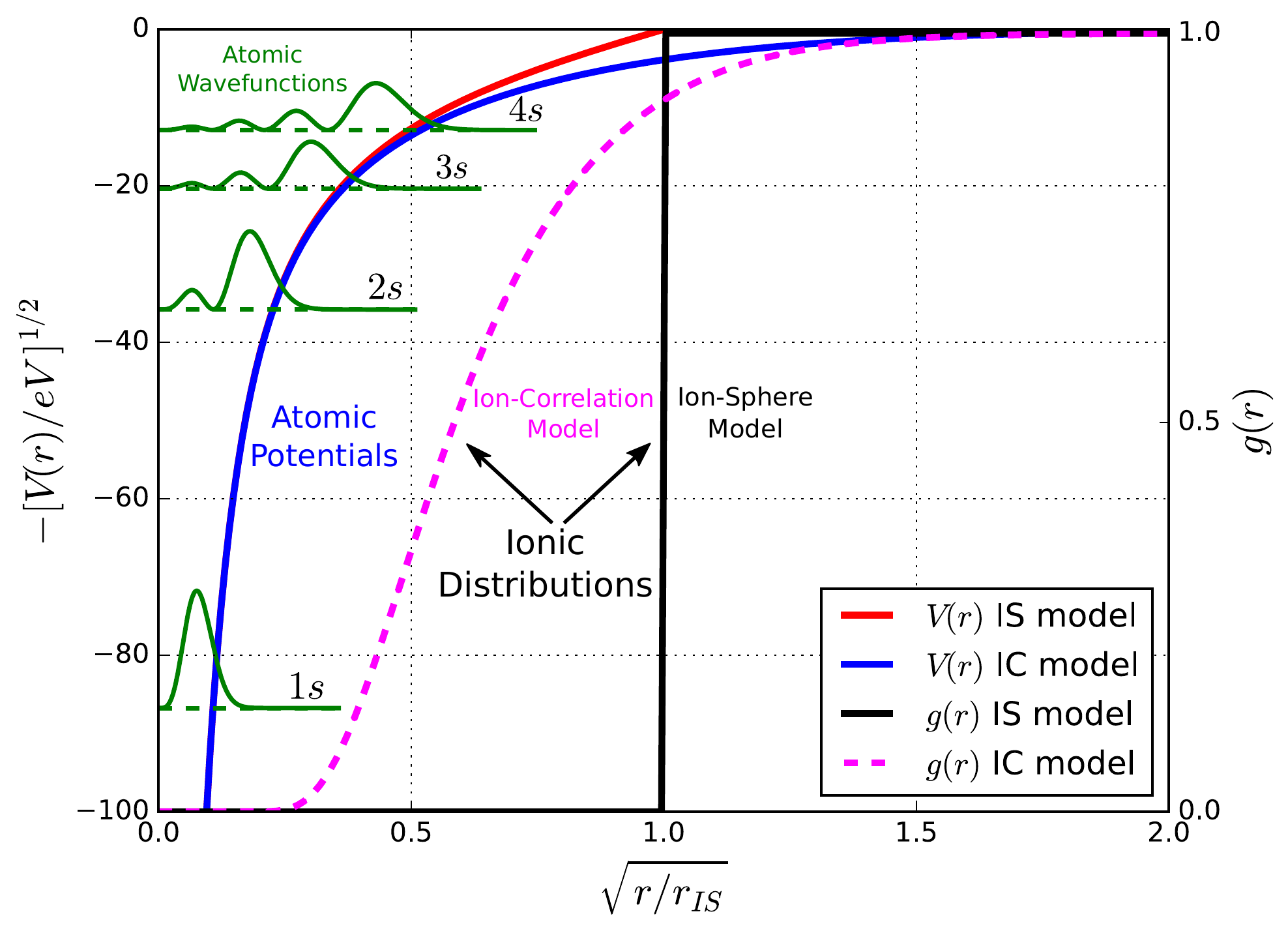}}
	\caption{
		\textbf{Models for the plasma surrounding an atom.}
		Shown are the central atomic potentials $ V(r) $  (left axis)
		in the IS (red) and IC (blue) model, as well as the ion-ion pair correlation functions $ g(r) $ (right axis) in the IS model (step function, in black) and IC model (magenta). The radial wavefunctions of the first four "s" orbitals are given in green (arbitrary units) where the vertical positions represent the orbital energies (on the left axis).
		The calculation was performed for iron near the convection zone boundary at $ R/R_{\odot}=0.72$, temperature $ T=180\text{eV} $ and electron density $ n_{e} =8\cdot 10^{22}\text{cm}^{-3}$.
	}
	\label{fig:pot_g}
\end{figure}

We have employed the opacity code STAR (\cite{krief2015effect,krief2015variance,krief2016solar,krief2016line}) to study the ion-correlation effects on the opacity of a 24-element solar mixture along a thermodynamic path obtained from a solar model implemented by Villante et al. (\cite{villante2014chemical}) which is calibrated with the recent AGSS09 set of chemical abundances (\cite{asplund2009chemical}). Bound-bound and bound-free atomic calculations are performed in the STA framework (\cite{BarShalom1989}) and based on a fully relativistic quantum mechanical theory via the Dirac equation, while inverse bremsstrahlung is calculated via a screened hydrogenic approximation with a degeneracy correction (\cite{krief2016solar}). Calculations were performed in the IS and IC approximation. In the IS model each ion is enclosed within a spherical volume whose radius $ r_{IS} $ is the Wigner-Seitz radius of the plasma. In addition, the ionic distribution vanishes inside the sphere while it is smeared out completely outside of it, so that the ion-ion pair correlation function $ g(r) $ takes the form of a step function and due to overall charge neutrality the central potential vanishes outside the sphere. Such a model neglects the possibility for ionic charge to penetrate into the ion sphere and screen the central potential seen by atomic wavefunctions, as illustrated in \autoref{fig:pot_g}. On the other hand, the IC model of Rozsnyai (\cite{rozsnyai1991photoabsorption,rozsnyai1992solar,rozsnyai2014equation}) includes a more realistic treatment of ionic structure which is solved self-consistently with the electronic structure and which screens the central potential. For strongly coupled plasmas (lower temperatures and higher densities) the ionic structure features the well known short range order, whereas for more weakly coupled plasmas, such as the solar mixture, it is smeared out while ions penetrate within the ion-sphere, as described in \autoref{fig:pot_g}. The longer range of the central potential in the IC model slightly widens the spatial range of bound orbitals. This causes an increase of the spatial overlap between bound and continuum states which in turn increase the photoionization cross section, enabling a non-negligible increase of the Rosseland opacity. In addition, the longer range of the potential slightly lowers the average ionization. We note that solar opacity calculations in the IC framework were performed in the past by \cite{rozsnyai1992solar}, using the older solar abundances of \cite{grevesse1993cosmic}.
Since our goal in this work is to estimate sensitivities, the IC calculations were performed in the Debye-Huckel approximation (see Equations 2.16-2.8 in \cite{rozsnyai1992solar}). This approximation results from the more accurate HNC approximation (which was used by \cite{rozsnyai1992solar}) in the limit of low plasma coupling.

\begin{figure}[]
	\resizebox{0.5\textwidth}{!}{\includegraphics{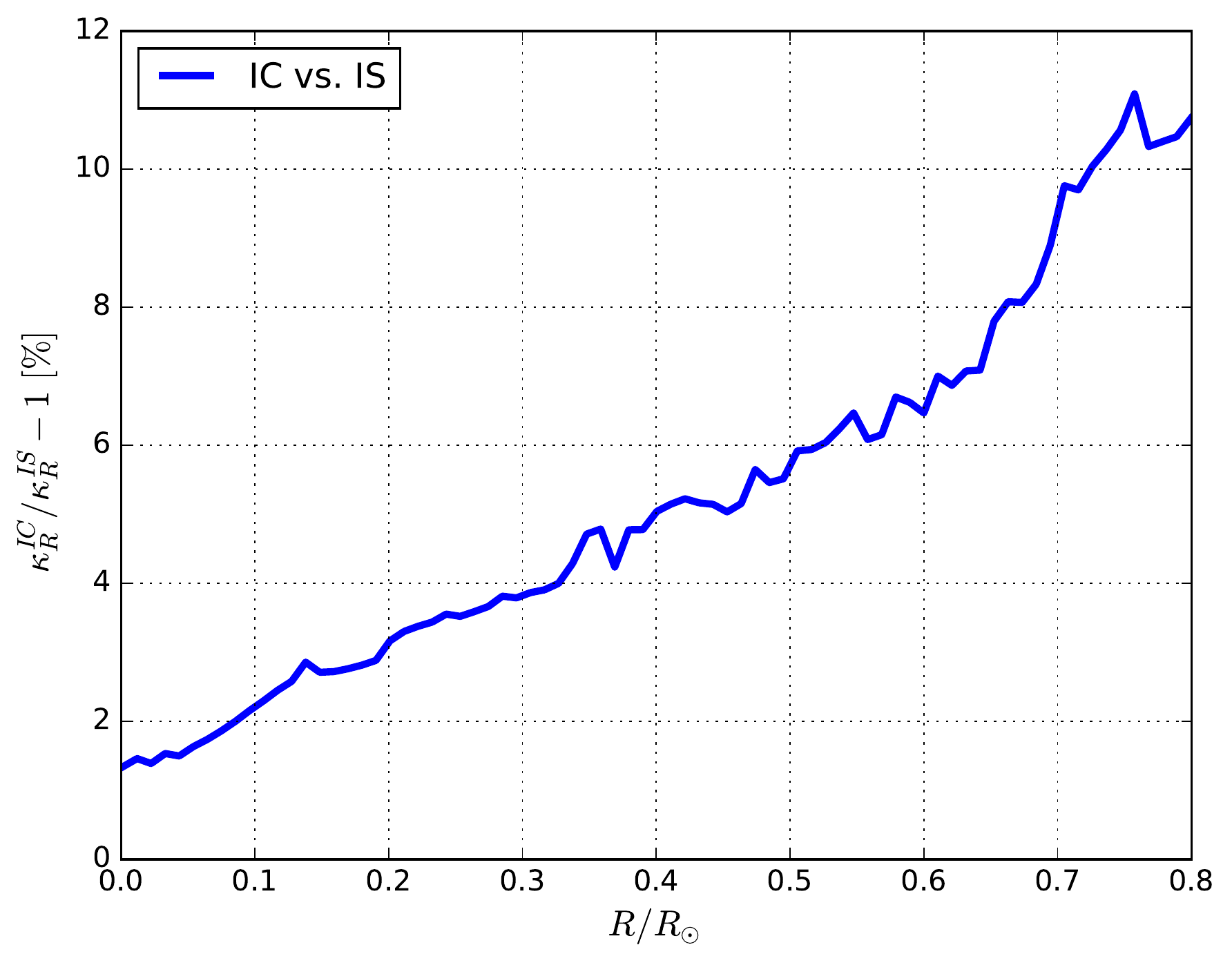}}
	\caption{
		Relative differences in the Rosseland opacity of the solar mixture throughout the solar interior, between Ion-Sphere and Ion-Correlation models calculated by STAR.
	}
	\label{fig:dk_ic}
\end{figure}

\begin{figure}[]
	\resizebox{0.5\textwidth}{!}{\includegraphics{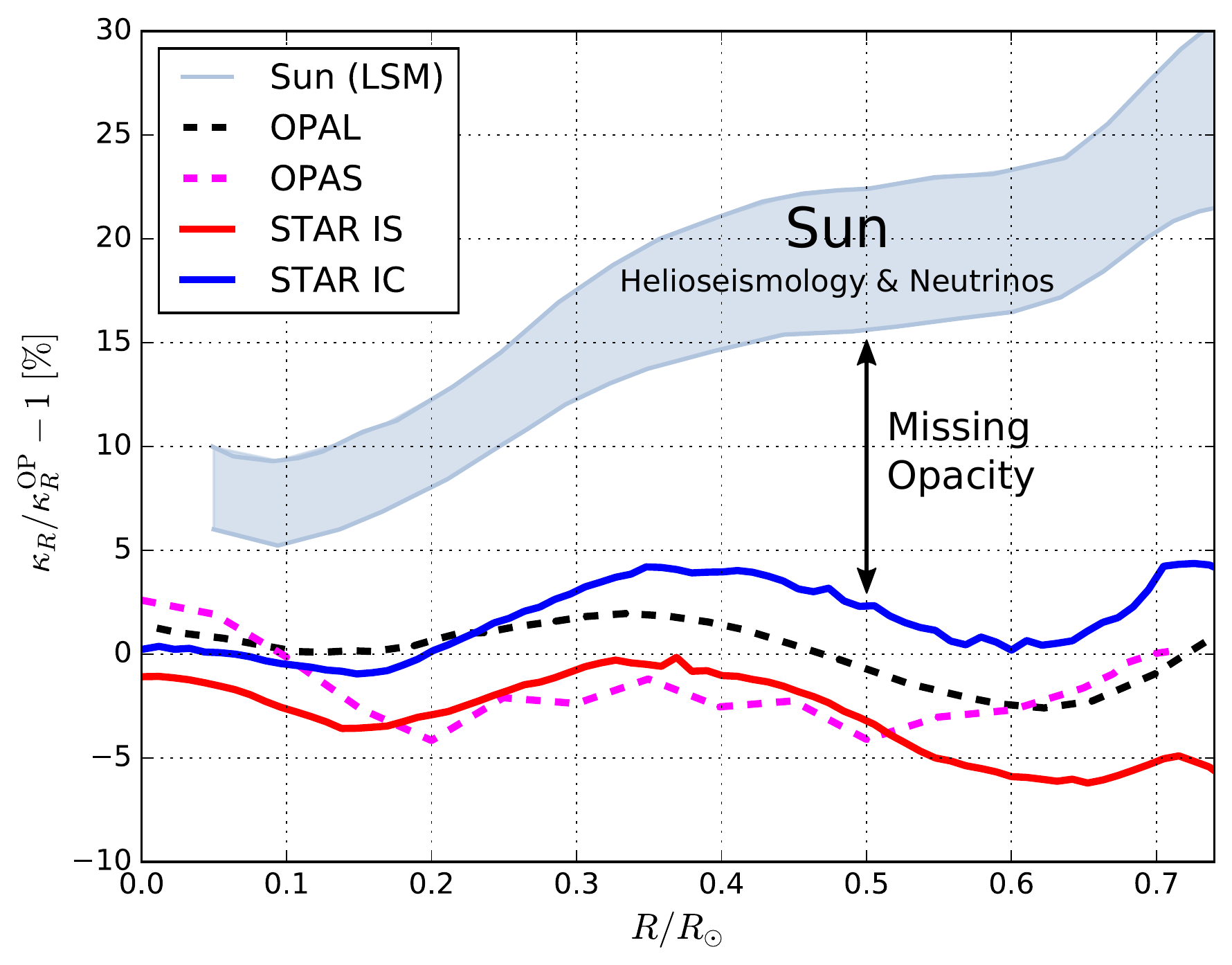}}
	\caption{
		The effect of plasma correlations on the solar opacity, as calculated by STAR,  relative to the OP. The differences with other models are also shown (as in \autoref{fig:dk_intro}).
	}
	\label{fig:dk_ic_op}
\end{figure}

The resulting variation of the Rosseland opacity of the solar mixture throughout the solar interior, between IC and IS as calculated by STAR, is shown in \autoref{fig:dk_ic}. 
These opacity variations are also shown in \autoref{fig:dk_ic_op}, relative to the OP.


\section{The effect of ion correlation in the Sandia Z-pinch experiment}
\begin{figure}[]
	\resizebox{0.5\textwidth}{!}{\includegraphics{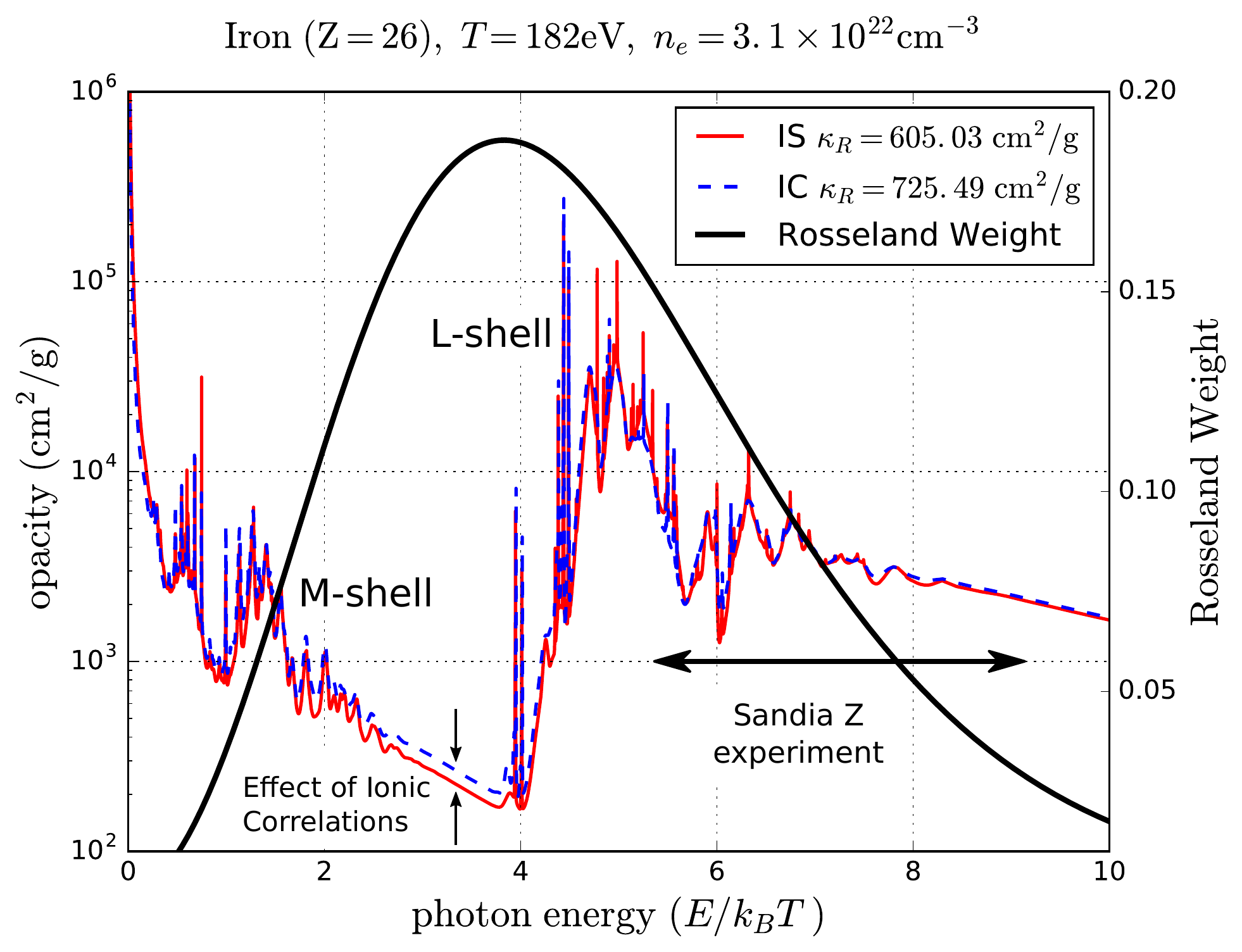}}
	\caption{
				\textbf{The effect of plasma correlations in the recent Sandia experiment.} The spectral opacity of iron (right axis) calculated by the IS (solid line, in red) and IC models (dashed line, in blue) at typical conditions reached by the recent Z-machine experiment (\cite{bailey2015higher})
				(temperature $ T=182\text{eV} $, free electron density $ n_{e}=3.1\times 10^{22}\text{cm}^{-3} $). The Rosseland weight function (right axis) is shown in the solid black line.  The experimentally accessible spectral range is indicated by a horizontal arrow. The Rosseland mean opacities are given in the legend.
	}
	\label{fig:specs_bailey}
\end{figure}

In \autoref{fig:specs_bailey}, the monochromatic opacities of iron calculated by STAR, at the conditions reached in the recent Sandia experiment (\cite{bailey2015higher})  are shown. Similar results were obtained by using the CRSTA atomic code (\cite{hazak2012configurationally,kurzweil2013inclusion}). The resulting Rosseland opacity in the IC model is larger by $ \sim 15\% $ than in the IS model, due to a significant increase of the M-shell photoionization cross-section, which results from the larger overlap of the M-shell bound orbital with the continuum states in the IC calculation. We note that despite having a significant contribution to the Rosseland opacity, which peaks at a photon energy of $ 3.8k_{B}T $, the M-shell is currently not accessible in the $ 5.5-9.5k_{B}T $ spectral range reached in opacity experiments at the Sandia Z machine, which includes only the L-shell spectra.
We  note that such required soft X-ray spectrometers are currently being developed (\cite{benstead2016new}).
It is seen from \autoref{fig:specs_bailey} that the plasma correlation has a smaller effect on the more strongly bound L-shell orbitals, and therefore cannot explain the "higher than predicted" iron opacity (\cite{bailey2015higher}), which is found in the L-shell spectral range. Hence, the correlation effect for iron cannot be measured currently in such a monochromatic experiment.

\section{Measurements of Rosseland opacities in radiation flow experiments}
The aforementioned monochromatic Z experiments give detailed information on opacity spectra, but have several drawbacks: (i) the inability to access the more important spectral range closer to the Rosseland peak, (ii) only electron densities smaller by a factor of $ \sim 2.5 $ than those  found near the CZB can be reached due to hydrodynamic expansion of the sample (\cite{nagayama2016calibrated}), (iii) the measured electron densities in such experiments are uncertain by  $ 20-30\% $ (\cite{nagayama2016model}) since they are inferred from crude line broadening models for Mg K-shell lines, and (iv) higher temperatures found in deeper regions in the Sun cannot be reached.

\begin{figure}[]
	\resizebox{0.5\textwidth}{!}{\includegraphics{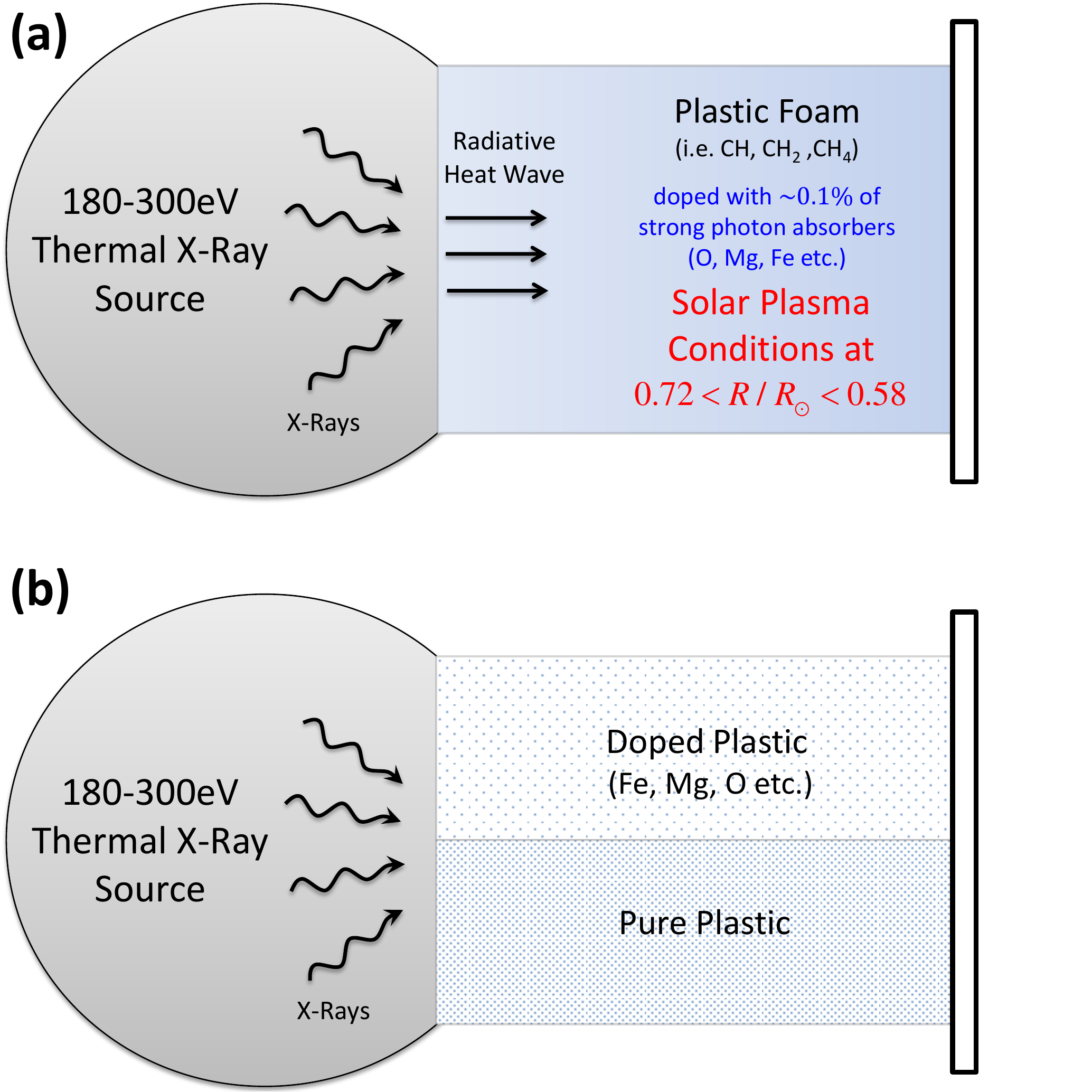}}
		\resizebox{0.5\textwidth}{!}{\includegraphics{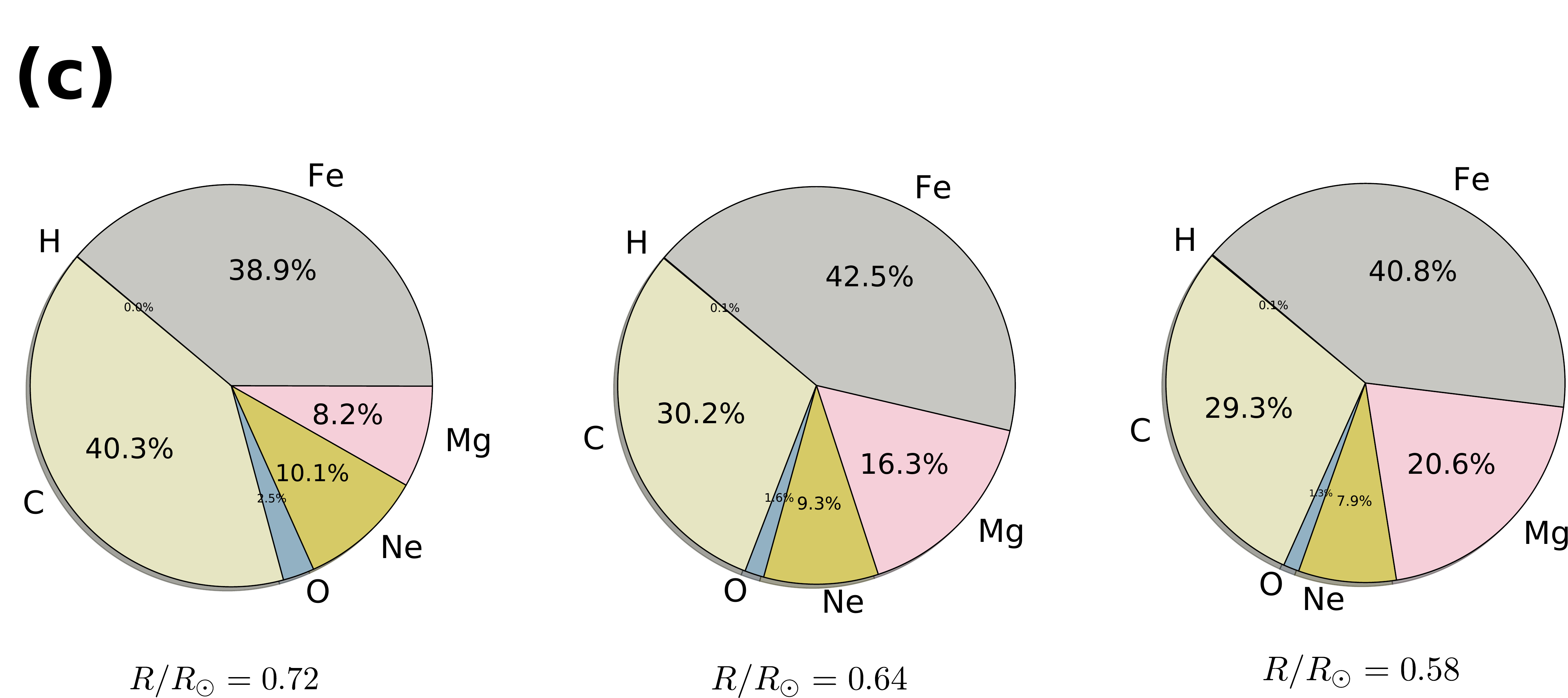}}
	\caption{\textbf{Illustration of a terrestrial radiation flow experiment to simulate the hot dense solar plasma.} (\textbf{a}) The side view shows a thermal x-ray source, coupled to a plastic foam such as CH, CH$ _{2} $ and  CH$ _{4} $. An x-ray drive with a typical  temperature of $ 180-300 $eV, which can be produced at high energy density facilities, generates a supersonic radiative heat wave through the foam. The Rosseland opacity of the foam is inferred directly from the measured breakout time of the radiative wave through the foam. By doping the foam with permilles of strong photon absorbers in the Sun, such as oxygen and iron, the Rosseland opacity of a mixture with plasma parameters identical to those prevailing in the solar interior can be studied experimentally.
	(\textbf{b}) Same as (a), with a foam setup consisting pure and doped plastic foams. In this setup, Rosseland opacity fractions can be measured directly, while the experimental error is reduced significantly, since both foams see the same x-ray drive.
	(\textbf{c}) The charts represent Rosseland opacity fractions of the CH  foam due to hydrogen, carbon and the doped elements, O, Ne, Mg, and Fe (with doping of $ 0.25\% $ each, see text) at three thermodynamic conditions representing  different positions in the Sun:
	$R/R_{\odot}=0.72 $ (left panel, $ T=180 $eV, $ n_{e}=8\times 10 ^{22}\text{cm}^{-3} $), $R/R_{\odot}=0.64 $ (middle panel, $ T=235 $eV, $ n_{e}=1.7\times 10 ^{23}\text{cm}^{-3} $) and $R/R_{\odot}=0.58 $ (right panel, $ T=280 $eV, $ n_{e}=3.1\times 10 ^{23}\text{cm}^{-3} $). Additional properties of the CH foam under these conditions are listed in \autoref{tab:expr}
	}
	\label{fig:expr}
\end{figure}

A complementary approach to the direct spectral measurements of solar opacities is via radiation flow experiments (\cite{back2000diffusive,fryer2016uncertainties}) in high energy density facilities, 
such as the National Ignition Facility (NIF) (\cite{moore2015characterization}), Laser-Megajoule (\cite{fleurot2005laser}, OMEGA (\cite{decker1997hohlraum}) and Sandia Z-machine (\cite{bailey2015higher}). In such experiments,  illustrated in \autoref{fig:expr}a, a high temperature thermal x-ray source is coupled to a foam, inducing a supersonic radiative x-ray wave crossing that foam. A common measurement is of the breakout time. This is an integral quantity which depends upon a combination of the equation of state of the foam used in the experiment and its opacity. 
Doping the foam with permilles of dominant photon absorbers in the Sun, such as oxygen, magnesium and iron, and comparing the measurements to those of a clean foam experiment, the Rosseland opacity can be studied experimentally at solar interior conditions that were never achieved in the past. The solar radius probed in the experiment is dictated by the foam initial density and the temperature of the heat wave. For example, recent results at the NIF show that a drive temperature of up to $ \sim 300 $eV can be reached (\cite{moore2015characterization}), which 
covers the temperatures in the Sun ranging from the CZB ($ R/R_{\odot}=0.72 $) and deeper, up to $ R/R_{\odot}\approx 0.55 $. By using low-Z foams, electron densities prevailing in the latter region of the Sun can also be achieved. 

As an example, in \autoref{tab:expr} we list the properties of the CH, CH$_{2}  $ and CH$ _{4} $ plastic foams, doped with oxygen, neon, magnesium and iron with a number concentration of $ 0.25\%  $ each,
at conditions found in the Sun at four different locations: $ R/R_{\odot}=0.72,0.64,0.58 $ and $ 0.55 $. 
The foam densities corresponding to conditions at the solar interior  are in the range of $ 0.3-1.3 \text{g/cm}^{3} $. 
Plastic foams can be easily manufactured with various hydrogen to carbon ratios and with densities of up to $ \sim 1\text{g/cm}^{3} $, which corresponds to $ R/R_{\odot}=0.58 $, where the temperature is $ 280 $eV.

Thus, radiation flow experiments introduce a few advantages: (i) Only Rosseland opacities are measured, (ii) plastic foams can be manufactured at densities corresponding to the solar plasma in the range $ R/R_{\odot} =0.72-0.58$ and  there is no hydrodynamic expansion of the foam since the radiative heat wave is supersonic, (iii) the foam density, which is known to a high precision, determines the electron density and (iv)  temperatures of up to $ 300 $eV can be reached (\cite{moore2015characterization}). Moreover, since the suggested doping level is very low, the equation of state of the foam is unaffected by the doping, as the chemical potential is dictated by the foam constituents, thus a comparative study disentangles the effects of the equation of state and the opacity in the velocity of the heat wave in the foam.

However, such experiments are plagued with systematic effects, as recently explained in \cite{fryer2016uncertainties}. The schematic experiments proposed here are of comparative nature, since it is only the effect of the doping elements which needs to be characterized. This reduces the sensitivity of the inferred opacities on the main sources of uncertainties.  In particular, the characterization of the x-ray source, which is essential for inferring the temperature of the supersonic wave, is a source of uncertainty, as it is not measured directly, but inferred from measurements without the foam, implicitly assuming repeatability in the operation of the x-ray source, and relying on theory to take into account the effect of the foam.
For example, Rosseland opacity fractions of test elements can be easily measured by comparing the results of foams with and without the tested element. We note that calibration and comparison of measurements with different foam setups is easily achieved, as shown in \autoref{fig:expr}b, by aligning the foams together, so that they would experience the same radiation drive.

This approach can also be used to pinpoint the ion-correlation effect, as it amounts to $ 5-10 \% $ in the Rosseland opacity (see \autoref{tab:expr}), and therefore, can be validated experimentally (\cite{fryer2016uncertainties}). 
The small amount of doping does not affect thermodynamic properties of the plasma such as the chemical potential, average ionization and electron density, while it significantly affects radiative properties. In \autoref{fig:expr}c we list the contributions of the different elements in the doped CH foam to its Rosseland opacity. It is evident that the majority of the opacity ($ \gtrsim 60\%$) is due to the doped elements, with iron having the largest contribution ($ \sim 40 \% $), similar to the solar plasma.

Hence, radiation flow experiments consist an ideal method for an experimental study of Rosseland opacities of the solar plasma, complementary and advantageous to monochromatic Z experiments. The experiments described here schematically should
be planned in detail using a dynamic code, for the specific machine. We postpone this
to a different publication.

\begin{table}[]
	\centering
	\caption{\textbf{Properties of doped plastic foams at terrestrial radiation flow experiments at solar interior conditions}. The average ionization $ \bar{Z} $, required mass density $ \rho $, Rosseland mean free path $ l_{R} $ and the relative difference in the Rosseland opacity between IC and IS, $ \Delta\kappa _{R}$, are listed for doped  CH, CH$ _{2} $ and CH$ _{4} $ (see text), with thermodynamic conditions (temperature $ T $ and electron density $ n_{e} $), prevailing at the listed locations $ R/R_{\odot} $ in the Sun. At $ R/R_{\odot}=0.58 $ the required foam densities approach the maximal available density of $ \sim 1 \text{g/cm}^{3}$  and at $ R/R_{\odot}=0.55 $ the temperature of $ 300 $eV is the maximal temperature which is currently achievable in terrestrial laboratories.}
	
	\label{tab:expr}
	\begin{tabular}{clccc}
		\hline & & CH  & CH$ _{2} $& CH$ _{4} $
		\\ \hline
		\multicolumn{1}{c|}{\begin{tabular}[c]{@{}c@{}}$ R/R_{\odot}=0.72 $\\ $ T=180 $eV\\ $ n_{e}=8\times 10 ^{22}\text{cm}^{-3} $\end{tabular}} & \begin{tabular}[c]{@{}l@{}}
			$ \bar{Z}  $\\ $ \rho\left[ \text{g/cm}^{3}\right]  $\\  $ l_{R}\left[ \text{mm}\right]  $\\ $ \Delta \kappa_{R}(\text{IC/IS}) $ 
		\end{tabular} 
		& \begin{tabular}[c]{@{}c@{}} 3.25 \\ 0.27\\ 0.13\\ 6.8$ \% $\end{tabular} & \begin{tabular}[c]{@{}c@{}}2.53\\ 0.26\\ 0.12\\ 7.2$ \% $\end{tabular} & \begin{tabular}[c]{@{}c@{}}1.94\\ 0.24\\ 0.12\\ 10.2$ \% $\end{tabular} \\ \hline
		\multicolumn{1}{c|}{\begin{tabular}[c]{@{}c@{}}$ R/R_{\odot}=0.64 $\\ $ T=235 $eV\\ $ n_{e}=1.7\times 10 ^{23} \text{cm}^{-3} $\end{tabular}} & 
		\begin{tabular}[c]{@{}l@{}}$ \bar{Z}  $\\$ \rho\left[ \text{g/cm}^{3}\right]  $\\  $ l_{R}\left[ \text{mm}\right]  $\\ $ \Delta \kappa_{R}(\text{IC/IS}) $ \end{tabular}
		& \begin{tabular}[c]{@{}c@{}}3.29\\ 0.58\\ 0.12\\ 5.1$ \% $\end{tabular} & \begin{tabular}[c]{@{}c@{}}2.55\\ 0.54\\ 0.11\\ 6.6$ \% $\end{tabular} & \begin{tabular}[c]{@{}c@{}}1.96\\ 0.5\\ 0.1\\ 7.1$ \% $\end{tabular}
		\\ \hline
		\multicolumn{1}{c|}{\begin{tabular}[c]{@{}c@{}}$ R/R_{\odot}=0.58 $\\ $ T=280 $eV\\ $ n_{e}=3.1\times 10 ^{23} \text{cm}^{-3} $\end{tabular}} & 
		\begin{tabular}[c]{@{}l@{}}$ \bar{Z}  $\\$ \rho\left[ \text{g/cm}^{3}\right]  $\\  $ l_{R}\left[ \text{mm}\right]  $\\ $ \Delta \kappa_{R}(\text{IC/IS}) $ \end{tabular}
		& \begin{tabular}[c]{@{}c@{}}3.29\\ 1.05\\ 0.1\\ 4.8$ \% $\end{tabular} & \begin{tabular}[c]{@{}c@{}}2.55\\ 0.99\\ 0.09\\ 4.9$ \% $\end{tabular} & \begin{tabular}[c]{@{}c@{}}1.96\\ 0.91\\ 0.08\\ 7.2$ \% $\end{tabular}
		\\ \hline
		\multicolumn{1}{c|}{\begin{tabular}[c]{@{}c@{}}$ R/R_{\odot}=0.55 $\\ $ T=300 $eV\\ $ n_{e}=4.2\times 10 ^{23} \text{cm}^{-3} $\end{tabular}} & 
		\begin{tabular}[c]{@{}l@{}}$ \bar{Z}  $\\$ \rho\left[ \text{g/cm}^{3}\right]  $\\  $ l_{R}\left[ \text{mm}\right]  $\\ $ \Delta \kappa_{R}(\text{IC/IS}) $ \end{tabular}
		& \begin{tabular}[c]{@{}c@{}}3.29\\ 1.42\\ 0.09\\ 5.3$ \% $\end{tabular} & \begin{tabular}[c]{@{}c@{}}2.55\\ 1.34\\ 0.08\\ 5.6$ \% $\end{tabular} & \begin{tabular}[c]{@{}c@{}}1.96\\ 1.23\\ 0.07\\ 5.8$ \% $\end{tabular}
		\\ \hline
	\end{tabular}
\end{table}

\section{Summary}
A significant increase to solar opacity calculations is required to solve the decade old disagreement between helioseismology and solar models. The small \textit{discrepancy} in the solar mixture opacity between all existing atomic models is not a measure of the \textit{accuracy} of these calculations, since similar physical models and approximations are used. An experimental validation of opacity calculations at stellar interior conditions, which has never been performed in the past, should aid significantly in understanding the source of the missing opacity.
 
Following the work of 
\cite{rozsnyai1991photoabsorption,rozsnyai1992solar},
the effect of ion-ion correlations which results from the extremely complex interaction between ions in a plasma, was examined as a source of uncertainty in the calculation of solar opacities.
 The effect was studied by using the Ion-Correlation model 
 which includes a more realistic treatment of ionic structure and was not previously analyzed, in the context of the solar opacity problem. We found a significant difference between the Ion-Sphere and Ion-Correlation models of $ \sim 10\% $ in the Rosseland opacity near the CZB. Both these models include uncontrolled approximations, and thus the difference between them should be considered as a measure of the uncertainty stemming in the microscopic ionic structure of the solar plasma.
This calls for a detailed examination of the effect with more advanced plasma models (\cite{daligault2016ionic}). 
In a recent work (\cite{krief2016line}) we have shown that uncertainties in line-broadening models, another plasma effect, have a significant contribution to the missing opacity, and in fact, can account for all of the missing opacity, if the line-width uncertainty is of a factor of about $ \sim 100 $. The uncertainties in line broadening models which are used in opacity codes are unknown, as they were never tested experimentally at stellar interior conditions. We note that differences between line-width models are typically factors of $ 2-50$. Changing artificially the width of the lines enabled a broad study of the effect of this phenomenon, in contrast to ionic-correlations, where we to rely upon the differences between specific models to estimate uncertainties. Thus, a
 detailed analysis which includes a sound theoretical or experimental quantification of the uncertainties due the various plasma effects, which depends solely on the temperature and free electron density, is called for.


We thus suggest an experimental path to study Rosseland opacities of the solar plasma via radiation flow experiments,
by using plastic foams doped with permilles of dominant photon absorbers in the Sun. We have shown that it is possible to cover the temperatures and densities of the solar plasma at solar radii  ranging from the CZB ($ R/R_{\odot}=0.72 $) and deeper, up to $ R/R_{\odot}\approx 0.58 $ which were never achieved in laboratory experiments. The analysis we have presented shows that doping materials contribute about half of the Rosseland opacity of the foam, even when the doping level is below $1\%$. Thus, we suggested that comparative measurements of clean and doped foam, with changing doping levels and various thermodynamic conditions, can aid in pin-pointing the opacity contribution of the different doping materials. Ion-correlations have a 5-10\% effect in the suggested setups, depending on the average ionization in the foam. By changing the carbon to hydrogen ratio one can try to isolate this effect in the experiments. A detailed analysis of these suggested experiments should include radiative-hydrodynamic simulation, to study possible systematic effects, such as wave front structure originating in boundary effects.

Finally, we stress out that by comparing the results of such experiments with theoretical predictions, the uncertainty in Rosseland opacity calculations at stellar interior conditions can be inferred for the first time.  If experimental results at the CZB conditions will deviate from calculations by more than $ 10-20\% $, the solar abundance problem will be solved, as the uncertainty in opacity calculations will be large enough for solar models to agree with helioseismology. Otherwise, it will be proven that the problem with solar models is not due to the opacities.



\begin{thebibliography}{}
	\expandafter\ifx\csname natexlab\endcsname\relax\def\natexlab#1{#1}\fi
	
	\bibitem[{Asplund {et~al.}(2009)Asplund, Grevesse, Sauval, \&
		Scott}]{asplund2009chemical}
	Asplund, M., Grevesse, N., Sauval, A.~J., \& Scott, P. 2009, arXiv preprint
	arXiv:0909.0948
	
	\bibitem[{Back {et~al.}(2000)Back, Bauer, Hammer, Lasinski, Turner, Rambo,
		Landen, Suter, Rosen, \& Hsing}]{back2000diffusive}
	Back, C., Bauer, J., Hammer, J., {et~al.} 2000, Physics of Plasmas
	(1994-present), 7, 2126
	
	\bibitem[{Bailey {et~al.}(2015)Bailey, Nagayama, Loisel, Rochau, Blancard,
		Colgan, Cosse, Faussurier, Fontes, Gilleron, {et~al.}}]{bailey2015higher}
	Bailey, J., Nagayama, T., Loisel, G., {et~al.} 2015, Nature, 517, 56
	
	\bibitem[{Bar-Shalom {et~al.}(1989)Bar-Shalom, Oreg, Goldstein, Shvarts, \&
		Zigler}]{BarShalom1989}
	Bar-Shalom, A., Oreg, J., Goldstein, W., Shvarts, D., \& Zigler, A. 1989,
	Physical Review A, 40, 3183
	
	\bibitem[{Benstead {et~al.}(2016)Benstead, Moore, Ahmed, Morton, Guymer,
		Soufli, Pardini, Hibbard, Bailey, Bell, {et~al.}}]{benstead2016new}
	Benstead, J., Moore, A., Ahmed, M., {et~al.} 2016, Review of Scientific
	Instruments, 87, 055110
	
	\bibitem[{Bergemann \& Serenelli(2014)}]{Bergemann2014}
	Bergemann, M., \& Serenelli, A. 2014, in Determination of Atmospheric
	Parameters of B-, A-, F-and G-Type Stars (Springer), 245--258
	
	\bibitem[{Blancard {et~al.}(2012)Blancard, Coss{\'e}, \&
		Faussurier}]{blancard2012solar}
	Blancard, C., Coss{\'e}, P., \& Faussurier, G. 2012, The Astrophysical Journal,
	745, 10
	
	\bibitem[{Blancard {et~al.}(2016)Blancard, Colgan, Coss{\'e}, Faussurier,
		Fontes, Gilleron, Golovkin, Hansen, Iglesias, Kilcrease,
		{et~al.}}]{blancard2016comment}
	Blancard, C., Colgan, J., Coss{\'e}, P., {et~al.} 2016, Physical Review
	Letters, 117, 249501
	
	\bibitem[{Christensen-Dalsgaard {et~al.}(2009)Christensen-Dalsgaard, Di~Mauro,
		Houdek, \& Pijpers}]{christensen2009opacity}
	Christensen-Dalsgaard, J., Di~Mauro, M.~P., Houdek, G., \& Pijpers, F. 2009,
	Astronomy \& Astrophysics, 494, 205
	
	\bibitem[{Ciricosta {et~al.}(2016)Ciricosta, Vinko, Barbrel, Rackstraw,
		Preston, Burian, Chalupsk{\`y}, Cho, Chung, Dakovski,
		{et~al.}}]{ciricosta2016measurements}
	Ciricosta, O., Vinko, S., Barbrel, B., {et~al.} 2016, Nature communications, 7
	
	\bibitem[{Colgan {et~al.}(2016)Colgan, Kilcrease, Magee, Sherrill, Jr., Hakel,
		Fontes, Guzik, \& Mussack}]{colgan2016new}
	Colgan, J., Kilcrease, D.~P., Magee, N.~H., {et~al.} 2016, The Astrophysical
	Journal, 817, 116
	
	\bibitem[{Daligault {et~al.}(2016)Daligault, Baalrud, Starrett, Saumon, \&
		Sjostrom}]{daligault2016ionic}
	Daligault, J., Baalrud, S.~D., Starrett, C.~E., Saumon, D., \& Sjostrom, T.
	2016, Physical review letters, 116, 075002
	
	\bibitem[{Decker {et~al.}(1997)Decker, Turner, Landen, Suter, Amendt, Kornblum,
		Hammel, Murphy, Wallace, Delamater, {et~al.}}]{decker1997hohlraum}
	Decker, C., Turner, R., Landen, O., {et~al.} 1997, Physical review letters, 79,
	1491
	
	\bibitem[{Dimitrijevic \& Konjevic(1987)}]{dimitrijevic1987simple}
	Dimitrijevic, M.~S., \& Konjevic, N. 1987, Astronomy and Astrophysics, 172, 345
	
	\bibitem[{Fleurot {et~al.}(2005)Fleurot, Cavailler, \&
		Bourgade}]{fleurot2005laser}
	Fleurot, N., Cavailler, C., \& Bourgade, J. 2005, Fusion Engineering and
	Design, 74, 147
	
	\bibitem[{Fryer {et~al.}(2016)Fryer, Dodd, Even, Fontes, Greeff, Hungerford,
		Kline, Mussack, Tregillis, Workman, {et~al.}}]{fryer2016uncertainties}
	Fryer, C., Dodd, E., Even, W., {et~al.} 2016, High Energy Density Physics, 18,
	45
	
	\bibitem[{Grevesse \& Noels(1993)}]{grevesse1993cosmic}
	Grevesse, N., \& Noels, A. 1993, in Origin and Evolution of the Elements,
	Vol.~1, 15--25
	
	\bibitem[{Haxton \& Serenelli(2008)}]{haxton2008cn}
	Haxton, W., \& Serenelli, A. 2008, The Astrophysical Journal, 687, 678
	
	\bibitem[{Hazak \& Kurzweil(2012)}]{hazak2012configurationally}
	Hazak, G., \& Kurzweil, Y. 2012, High Energy Density Physics, 8, 290
	
	\bibitem[{Iglesias(2015)}]{iglesias2015iron}
	Iglesias, C.~A. 2015, Monthly Notices of the Royal Astronomical Society, 450, 2
	
	\bibitem[{Iglesias \& Rogers(1996)}]{iglesias1996updated}
	Iglesias, C.~A., \& Rogers, F.~J. 1996, The Astrophysical Journal, 464, 943
	
	\bibitem[{Krief \& Feigel(2015{\natexlab{a}})}]{krief2015effect}
	Krief, M., \& Feigel, A. 2015{\natexlab{a}}, High Energy Density Physics, 15,
	59
	
	\bibitem[{Krief \& Feigel(2015{\natexlab{b}})}]{krief2015variance}
	---. 2015{\natexlab{b}}, High Energy Density Physics, 17, Part B, 254
	
	\bibitem[{Krief {et~al.}(2016{\natexlab{a}})Krief, Feigel, \&
		Gazit}]{krief2016line}
	Krief, M., Feigel, A., \& Gazit, D. 2016{\natexlab{a}}, The Astrophysical
	Journal, 824, 98
	
	\bibitem[{Krief {et~al.}(2016{\natexlab{b}})Krief, Feigel, \&
		Gazit}]{krief2016solar}
	---. 2016{\natexlab{b}}, The Astrophysical Journal, 821, 45
	
	\bibitem[{Kurzweil \& Hazak(2013)}]{kurzweil2013inclusion}
	Kurzweil, Y., \& Hazak, G. 2013, High Energy Density Physics, 9, 548
	
	\bibitem[{Liberman(1979)}]{liberman1979self}
	Liberman, D.~A. 1979, Physical Review B, 20, 4981
	
	\bibitem[{Mondet {et~al.}(2015)Mondet, Blancard, Coss{\'e}, \&
		Faussurier}]{mondet2015opacity}
	Mondet, G., Blancard, C., Coss{\'e}, P., \& Faussurier, G. 2015, The
	Astrophysical Journal Supplement Series, 220, 2
	
	\bibitem[{Moore {et~al.}(2015)Moore, Guymer, Morton, Williams, Kline, Bazin,
		Bentley, Allan, Brent, Comley, {et~al.}}]{moore2015characterization}
	Moore, A.~S., Guymer, T.~M., Morton, J., {et~al.} 2015, Journal of Quantitative
	Spectroscopy and Radiative Transfer, 159, 19
	
	\bibitem[{Nagayama {et~al.}(2016{\natexlab{a}})Nagayama, Bailey, Loisel,
		Rochau, MacFarlane, \& Golovkin}]{nagayama2016calibrated}
	Nagayama, T., Bailey, J., Loisel, G., {et~al.} 2016{\natexlab{a}}, Physical
	Review E, 93, 023202
	
	\bibitem[{Nagayama {et~al.}(2016{\natexlab{b}})Nagayama, Bailey, Loisel,
		Rochau, Hansen, Blancard, Cosse, Faussurier, Gilleron, Pain,
		{et~al.}}]{nagayama2016model}
	---. 2016{\natexlab{b}}, High Energy Density Physics
	
	\bibitem[{Rozsnyai(1991)}]{rozsnyai1991photoabsorption}
	Rozsnyai, B.~F. 1991, Physical Review A, 43, 3035
	
	\bibitem[{Rozsnyai(1992)}]{rozsnyai1992solar}
	---. 1992, The Astrophysical Journal, 393, 409
	
	\bibitem[{Rozsnyai(2014)}]{rozsnyai2014equation}
	---. 2014, High Energy Density Physics, 10, 16
	
	\bibitem[{Schlattl {et~al.}(1999)Schlattl, Weiss, \&
		Raffelt}]{schlattl1999helioseismological}
	Schlattl, H., Weiss, A., \& Raffelt, G. 1999, Astroparticle Physics, 10, 353
	
	\bibitem[{Seaton {et~al.}(1994)Seaton, Yan, Mihalas, \&
		Pradhan}]{seaton1994opacities}
	Seaton, M., Yan, Y., Mihalas, D., \& Pradhan, A.~K. 1994, Monthly Notices of
	the Royal Astronomical Society, 266, 805
	
	\bibitem[{Serenelli(2016)}]{serenelli2016alive}
	Serenelli, A. 2016, The European Physical Journal A, 52, 1
	
	\bibitem[{Serenelli {et~al.}(2016)Serenelli, Scott, Villante, Vincent, Asplund,
		Basu, Grevesse, \& Pe{\~n}a-Garay}]{serenelli2016implications}
	Serenelli, A., Scott, P., Villante, F.~L., {et~al.} 2016, Monthly Notices of
	the Royal Astronomical Society, 463, 2
	
	\bibitem[{Serenelli {et~al.}(2009)Serenelli, Basu, Ferguson, \&
		Asplund}]{serenelli2009new}
	Serenelli, A.~M., Basu, S., Ferguson, J.~W., \& Asplund, M. 2009, The
	Astrophysical Journal Letters, 705, L123
	
	\bibitem[{Serenelli {et~al.}(2011)Serenelli, Haxton, \&
		Pe{\~n}a-Garay}]{serenelli2011solar}
	Serenelli, A.~M., Haxton, W.~C., \& Pe{\~n}a-Garay, C. 2011, The Astrophysical
	Journal, 743, 24
	
	\bibitem[{Villante {et~al.}(2014)Villante, Serenelli, Delahaye, \&
		Pinsonneault}]{villante2014chemical}
	Villante, F.~L., Serenelli, A.~M., Delahaye, F., \& Pinsonneault, M.~H. 2014,
	The Astrophysical Journal, 787, 13
	
	\bibitem[{Vinyoles {et~al.}(2016)Vinyoles, Serenelli, Villante, Basu,
		Bergstr{\"o}m, Gonzalez-Garcia, Maltoni, Pe{\~n}a-Garay, \&
		Song}]{vinyoles2016new}
	Vinyoles, N., Serenelli, A.~M., Villante, F.~L., {et~al.} 2016, arXiv preprint
	arXiv:1611.09867
	
	\bibitem[{Vinyoles {et~al.}(2017)Vinyoles, Serenelli, Villante, Basu,
		Bergstr{\"o}m, Gonzalez-Garcia, Maltoni, Pe{\~n}a-Garay, \&
		Song}]{vinyoles2017new}
	---. 2017, The Astrophysical Journal, 835, 202
	
\end{thebibliography}

\end{document}